\documentclass[smallextended]{svjour3}%
\pdfoutput=1
\usepackage{graphics,amsmath,amsfonts,amscd,revsymb,latexsym,
enumerate,multirow,epsfig}
\usepackage{amsmath}
\usepackage{fullpage}
\usepackage{amsfonts}
\usepackage{amssymb}
\usepackage{graphicx}%
\providecommand{\U}[1]{\protect\rule{.1in}{.1in}}

\smartqed
\journalname{Foundations of Physics}
\begin{document}

\title{Addressing the clumsiness loophole in a Leggett-Garg test of macrorealism}
\author{Mark M. Wilde \and Ari Mizel}\institute{Mark M. Wilde is
a postdoctoral fellow in the School of Computer Science, McGill University, Montreal, Quebec, Canada. Ari Mizel is with the Laboratory for Physical Sciences, 8050 Greenmead Drive, College Park,
Maryland, USA 20740.  (E-mail: mark.wilde@mcgill.ca; ari@arimizel.com).}
\date{Received: \today / Accepted: }
\maketitle

\begin{abstract}
The rise of quantum information theory has lent new relevance to experimental
tests for non-classicality, particularly in controversial cases such as
adiabatic quantum computing superconducting circuits. The Leggett-Garg
inequality is a \textquotedblleft Bell inequality in time\textquotedblright%
\ designed to indicate whether a single quantum system behaves in a
macrorealistic fashion.  Unfortunately, a violation of the inequality can only show that the system is {\it either} (i) non-macrorealistic {\it or} (ii) macrorealistic but subjected to a measurement technique that happens to disturb the system.  The ``clumsiness'' loophole (ii) provides reliable refuge for the stubborn macrorealist, who can invoke it to brand recent experimental and theoretical work on the Leggett-Garg test inconclusive. Here, we present a revised Leggett-Garg protocol that permits one to conclude that a system is {\it either} (i) non-macrorealistic {\it or} (ii) macrorealistic but with the property that two seemingly non-invasive measurements can somehow collude and strongly disturb the system.  By providing an explicit check of the invasiveness of the measurements,  the protocol replaces the clumsiness loophole with a significantly smaller ``collusion'' loophole.
\end{abstract}

\section{Introduction}

One of the hallmarks of the quantum theory is that it
defies our intuition about the world. Faced with experimental manifestations
of entanglement and wave-particle duality, one may recognize the need to
revise classical theories, but one does not readily accept the radical quantum
picture of nature. Bell's theorem \cite{bell64,B87} provided a clear
experimental protocol to test whether nature could obey a revised classical
theory at least satisfying the minimal postulates of local realism. The
experimental demonstration of a violation of Bell's inequality
\cite{PhysRevLett.49.1804} showed that even these minimal postulates must be abandoned.

Unfortunately, the Bell protocol can be experimentally demanding, especially for large quantum systems such as superconducting qubits, requiring two parts that can be entangled, spatially separated, and independently measured. If one wishes to look for signatures of quantum behavior in such systems, a non-local Bell test is usually impractical.  Leggett and Garg therefore framed a less experimentally demanding protocol that tests for
violation of the postulates of \textit{macrorealism} rather than local realism
\cite{PhysRevLett.54.857,Home1984159}. Refinements exist in the literature,
but the original postulates of macrorealism are as follows: a macroscopic system with two or more macroscopically distinct states available to it will at all times be in one or the other of these states (\textit{macroscopic realism per se}), and it is
possible, in principle, to determine the state of the system with arbitrary small perturbation to its subsequent dynamics (\textit{non-invasive measurability}).
Leggett and Garg derived an inequality similar to that of Bell
\cite{PhysRevLett.54.857} to bound the temporal correlations observable in a
macrorealistic theory. (Some have referred to the Leggett-Garg inequality as a
\textit{Bell inequality in time}). A violation of the inequality is supposed
to show that a system is not behaving macrorealistically. 


Despite
being more experimentally tractable than a Bell test for systems like superconducting qubits, any Leggett-Garg test suffers from a serious vulnerability in comparison to a Bell test.
Since the locality postulate of local realism
states that it is impossible to affect a spatially distant physical system, a
violation of Bell's inequality can unequivocally demonstrate a failure of
local realism.  On the other hand, macrorealism does not assert that it is
impossible to affect a physical system by measurement but merely
that it is possible for a sufficiently adroit measurement to avoid doing so.  Thus the Leggett-Garg test can show only that the system is {\it either} (i) non-macrorealistic {\it or} (ii) macrorealistic but subjected to a measurement technique that happens to disturb the system.  Rather than abandoning a cherished view of the nature of physical
reality, an adherent of macrorealism will likely attribute a
violation of the Leggett-Garg inequality to the ``clumsiness'' loophole (ii), which results from experimental limitations, rather than the radical finding (i), which topples the macrorealist's picture of nature.  Recent experimental Leggett-Garg tests \cite{XLZG09,GABLOWP09,PMNBVEK10} and theoretical
work \cite{PhysRevLett.101.090403,PhysRevLett.99.180403,RKM06,JKB06,WJ08} are thus
inconclusive in the eyes of a stubborn macrorealist.  Leggett and Garg acknowledge this loophole but maintain that clever measurement schemes, such as ideal negative-result measurements,
argue against the macrorealist's retort.

In this paper, we provide a more methodical and general means of addressing the
 clumsiness loophole.  Of course, it is impossible in principle to prove once and for all that a measurement device is non-invasive.  Even if the device were to pass a number of tests for non-invasiveness, one never knows whether some test exists which the device would fail (i.e.,~a scientific hypothesis like ``the measurement device is non-invasive'' can be falsified but cannot be proven true once and for all).  Furthermore, in our context, we would need to demonstrate that the device would be non-invasive if not for the fact that the quantum system being measured cannot be measured non-invasively.  Our approach is instead to frame the notion of an \textquotedblleft adroit measurement\textquotedblright\ and show how to demonstrate it experimentally.  We then devise an experiment that shows how to use only adroit measurements to violate the Leggett-Garg inequality. The result is a compelling protocol that addresses the clumsiness loophole, while having more modest experimental requirements (for systems like superconducting qubits) than the Bell inequality.  It allows one to conclude that a system under investigation is {\it either} (i) non-macrorealistic {\it or} (ii) macrorealistic but with the property that two adroit measurements can somehow collude and strongly disturb the system.  The protocol advances beyond the original Leggett-Garg test by providing an explicit way to check whether the measurement technique happens to disturb the system.  The clumsiness loophole is thereby closed, although a significantly smaller collusion loophole remains since the check is not exhaustive.

We structure this paper as follows. We first review the standard Leggett-Garg
inequality and then present an ideal variation of it that replaces the
clumsiness loophole with the smaller collusion loophole as described above. We show how increasing the number of
adroit\ measurements leads to a stronger violation of the Leggett-Garg
inequality. We then generalize the scenario and report how the inclusion of
noise leads to a practical trade-off between the number of adroit measurements
and the strength of dephasing noise.  Our conclusion includes remarks on the clumsiness loophole as it relates to recent weak-measurement versions of the Leggett-Garg inequality  \cite{RKM06,JKB06,WJ08}.

\section{Leggett-Garg Inequality for Qubits}
Suppose that we prepare a given
system in some specific initial state and measure a dichotomic observable $Q$
(with realizations $\pm1$) after delays $t_{i}$, where $i=1,2,3$. We repeat
and average the results to compute the two-time correlation
functions $C_{i,j}\equiv\left\langle Q_{i}  Q_{j}  \right\rangle $,
where we employ the shorthand $Q_{i}\equiv Q\left(  t_{i}\right) $. The Leggett-Garg
\cite{PhysRevLett.54.857}  inequality states
\begin{equation}
\mathcal{L}\equiv C_{1,2}+C_{2,3}+C_{1,3}+1\geq0, \label{eq:leggett-ineq}%
\end{equation}
for any choice of $t_{1}$, $t_{2}$, and $t_{3}$.  As a matter of elementary mathematics, this inequality is always satisfied if we measure three $\pm1$-valued numbers $Q_{1}$ then $Q_{2}$ then $Q_{3}$, compute $Q_{1}Q_{2}$, $Q_{2}Q_{3}$, and $Q_{1}Q_{3}=(Q_{1}Q_{2})(Q_{2}Q_{3})$, repeat and average to obtain $C_{i,j}$.  The possibility of a violation only arises if we compute $C_{1,3}$ using a separate series of experiments in which $Q_{1}$ and $Q_{3}$ are measured but not $Q_{2}$; one can emphasize this by attaching a $^{\prime}$ superscript to $C_{1,3} = \left\langle Q_{1}Q_{3}\right\rangle ^\prime$.  A violation of (\ref{eq:leggett-ineq}) can only occur if $\left\langle Q_{1}Q_{3}\right\rangle ^\prime \ne \left\langle Q_{1}Q_{3}\right\rangle = \left\langle (Q_{1}Q_{2}) (Q_{2}Q_{3})\right\rangle $.  This should only happen if the act of measuring $Q_{2}$ ``matters," which should not be the case for a macrorealistic system being measured non-invasively. 

Suppose, then, that an experimentalist finds a violation of the inequality. How
should this result be interpreted?  Perhaps the system is not macrorealistic -- a violation of the above inequality
occurs, for example, for a two-state quantum system initialized to a maximally
mixed state in a noiseless environment with vanishing system Hamiltonian, and $Q_1 = \sigma_z$, $Q_2 = (\sigma_y - \sigma_z)/\sqrt{2}$, $Q_3 = -\sigma_y$.  (Here,  $\sigma_{i}$
denotes a Pauli matrix.)  \footnote{We are assuming
a vanishing Hamiltonian for now in order
simplify the presentation. Section~\ref{sec:non-zero-Ham} details an equivalent, but experimentally more natural case, in which there is a non-trivial Hamiltonian $\omega \sigma_X$, the observables are all $Q(t_i) = \sigma_Z$, and the measurement times $t_i$ are chosen appropriately. }    For such a quantum system, the measurement of $Q_{2}$ ``matters" because it causes a collapse according to the axioms of quantum mechanics.  On the other hand, a violation could also occur for a
macrorealistic two-state system being measured in an invasive fashion because
of limitations in the experimental measurement technique; in this case the act of measuring $Q_{2}$ matters only because experimental practicalities have led to a "clumsy" measurement apparatus.  Here, we propose a protocol that
is designed to mitigate this stark uncertainty in interpretation.

\begin{figure}[ptb]
\begin{center}
\includegraphics[
natheight=2.879800in,
natwidth=12.739500in,
width=3.0399in
]{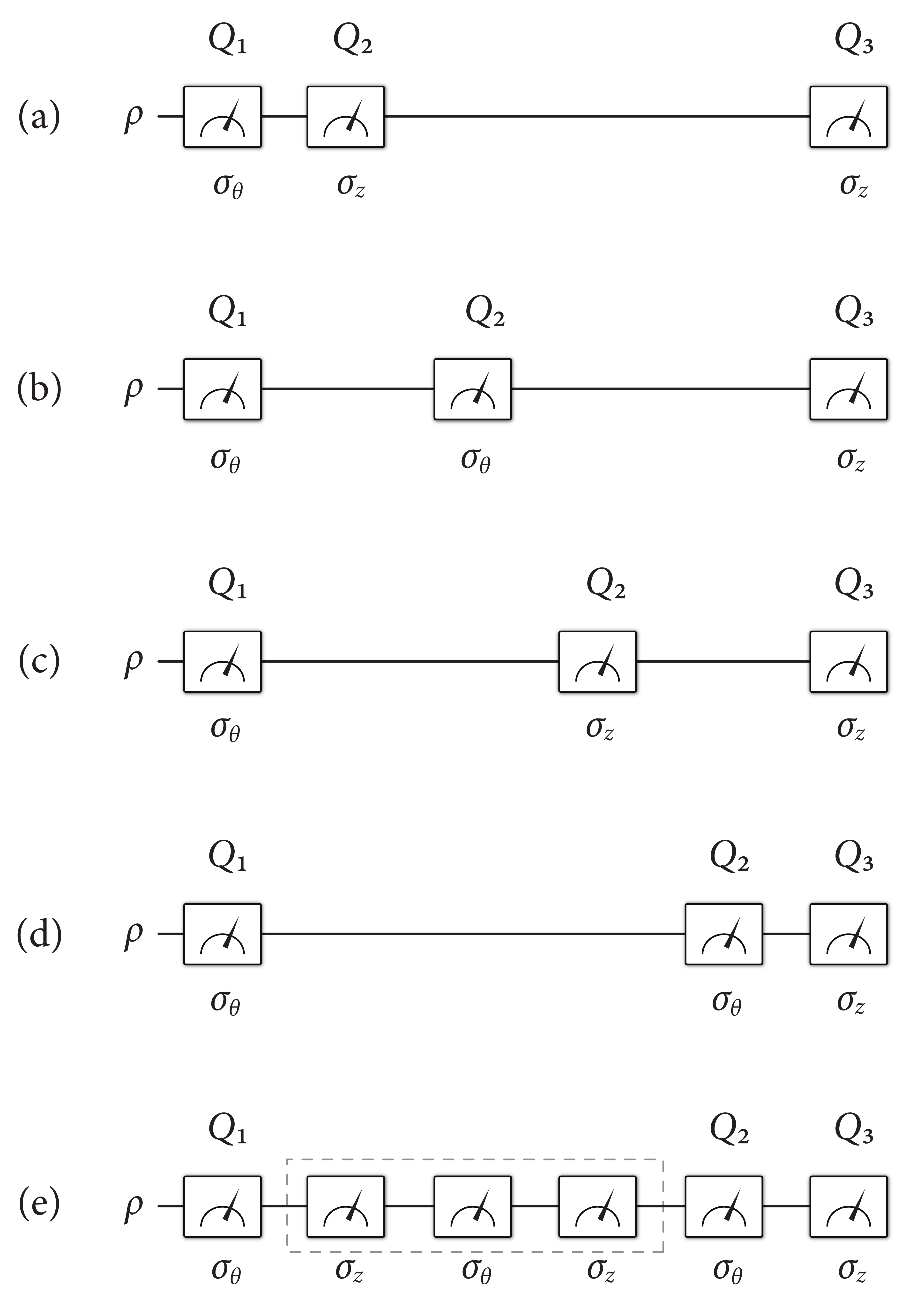}
\end{center}
\caption{A series of experiments to address the clumsiness loophole in a
violation of the Leggett-Garg inequality in equation (\ref{eq:leggett-ineq}). Suppose
that the second measurement in (a)-(d) is shown experimentally to have no
effect on the joint probability distribution between measurements of
observables $Q_{1}$ and $Q_{3}$. Then these measurements are
adroit, and any sequence of them is also
adroit. We then perform a new experiment (e) that
includes these measurements. Given the evidence of adroitness provided by
(a)-(d), plus the closure of adroitness axiom, violation of the Leggett-Garg inequality in (e) can be attributed to
a failure of macrorealism rather than to experimental clumsiness.}%
\label{leggett-experiment}%
\end{figure}

Consider the experiment depicted in Fig.~\ref{leggett-experiment}%
(a). We will say that the second measurement is adroit if it does not have any affect on the
joint probability distribution of the outcomes of the first and third
measurements. That is, suppose that results $a$ and $c$ correspond to
the respective outcomes of the first and third measurements. Then, the second
measurement is $\epsilon$-adroit if%
\begin{eqnarray*}
\sum_{a,c}&&\left\vert P\left\{  a,c\ |\text{2nd measurement is performed}\ \right\} \right.\\
 && \left.-P\left\{a,c\ |\ \text{2nd measurement not performed}\right\}  \right\vert \leq \epsilon,
\end{eqnarray*}
where we stipulate (as in Fig. 1a) that no other measurement be
performed besides the first, second, and third measurements.  An experimentalist can confirm whether a measurement is $\epsilon$-adroit (we will simply say \textquotedblleft adroit\textquotedblright\ in what follows) by collecting statistics to build confidence that the above condition holds.  Assuming that the $\epsilon$-adroitness test gives convincing evidence that the 2nd measurement has no effect on the system, it is natural to assume for a macrorealistic system that performing two $\epsilon$-adroit measurements should yield a $2\epsilon$-adroit composite measurement (this could be termed the \textit{closure of adroit measurements}).  Of course, if the 2nd measurement has some violent effect on the system that is not revealed in the $\epsilon$-adroitness test, it is possible in principle for two seemingly innocuous $\epsilon$-adroit measurements to collude and have a dramatic effect on the system, violating the closure of adroit measurements.  Our protocol does not rule out this (in our opinion unnatural) possibility.  

With this notion of adroit measurements in mind, consider the procedure depicted in Fig.~\ref{leggett-experiment}.  Suppose the experimentalist first performs experiments (a)-(d) and can demonstrate in
each case that the second measurement is adroit as defined above.  Finally, the
experimentalist performs Fig.~\ref{leggett-experiment}(e). Experiment (e) follows the
Leggett-Garg protocol except that $\left\langle Q_{1}Q_{2}\right\rangle $ and
$\left\langle Q_{2}Q_{3}\right\rangle $ are obtained in the presence of the
boxed measurements, whereas $\left\langle Q_{1}Q_{3}\right\rangle ^\prime$ is
obtained not only without performing $Q_{2}$, but also without performing the
boxed measurements.  Since the boxed measurements and $Q_2$ of Fig.~\ref{leggett-experiment}(e) have individually been shown to be adroit in experiments (a)-(d), the closure of adroit measurements implies that it should have negligible effect on the correlator $\left\langle Q_{1}Q_{3}\right\rangle ^\prime$ whether they are performed or not.


What happens if we apply this new procedure to a two-state system
such as a qubit that is actually quantum mechanical?  As shown in Fig.~\ref{leggett-experiment}, we
choose as observables $\sigma_{z}$ and $\sigma_{\theta}\equiv\cos
 \left( \theta\right)  \sigma_{z}+\sin \left( \theta \right) \sigma_{x}$.  Assume for now that
the system is noiseless and has vanishing Hamiltonian.  Then, it is clear
that the second measurement in experiments
(a)-(d) does not affect the correlations between the first and third
measurements because measuring a variable twice in a row has no additional effect on the system.  Thus, the measurements are perfectly adroit in this ideal case.  What happens in experiment (e)? To answer this question, we
first calculate several relevant quantities. We define superoperator
$\overline{\Delta}$ as a $\sigma_{z}$ basis dephasing of a qubit with density
operator $\rho$: $\overline{\Delta}\left(  \rho\right)  \equiv\frac{1}%
{2}\left(  \rho+\sigma_{z}\rho\sigma_{z}\right)  $ and $\overline{\Delta
}_{\theta}$ as a $\sigma_{\theta}$ basis dephasing: $\overline{\Delta}%
_{\theta}\left(  \rho\right)  \equiv\frac{1}{2}\left(  \rho+\sigma_{\theta
}\rho\sigma_{\theta}\right)  $. The following relation is useful: %
$
\overline{\Delta}\left(  \sigma_{\theta}\right)  =\cos\left(  \theta\right)
\sigma_{z}, \label{eq:dephasing}%
$
and a similar relation can be derived by exploiting it: %
$
\overline{\Delta}_{\theta}\left(  \sigma_{z}\right)   = \cos\left(  \theta\right) \sigma_{\theta}$ .
Now, we calculate the correlation 
$
\left\langle Q_{1}Q_{3}\right\rangle ^\prime =\frac{1}{2}\text{Tr}\left[  \sigma
_{z}\left\{  \sigma_{\theta},\rho\right\}  \right]  , \label{eq:q1-q3}%
$
where $\left\{  \sigma_{\theta},\rho\right\}  $ is the anticommutator. The
other correlation functions $\left\langle Q_{1}Q_{2}\right\rangle $ and
$\left\langle Q_{2}Q_{3}\right\rangle $\ are as follows:%
\begin{align}
\left\langle Q_{1}Q_{2}\right\rangle  &  =\frac{1}{2}\text{Tr}\left[
\sigma_{\theta}\left(  \overline{\Delta}\circ\overline{\Delta}_{\theta}%
\circ\overline{\Delta}\right)  \left(  \left\{  \sigma_{\theta},\rho\right\}
\right)  \right]  ,\label{eq:q1-q2}\\
\left\langle Q_{2}Q_{3}\right\rangle  &  =\frac{1}{2}\text{Tr}\left[
\sigma_{z}\left\{  \sigma_{\theta},\left(  \overline{\Delta}\circ
\overline{\Delta}_{\theta}\right)  ^{2}\left(  \rho\right)  \right\}  \right]
. \label{eq:q2-q3}%
\end{align}
%
\begin{figure}
[ptb]
\begin{center}
\includegraphics[
natheight=4.375100in,
natwidth=5.833200in,
width=3.4405in
]%
{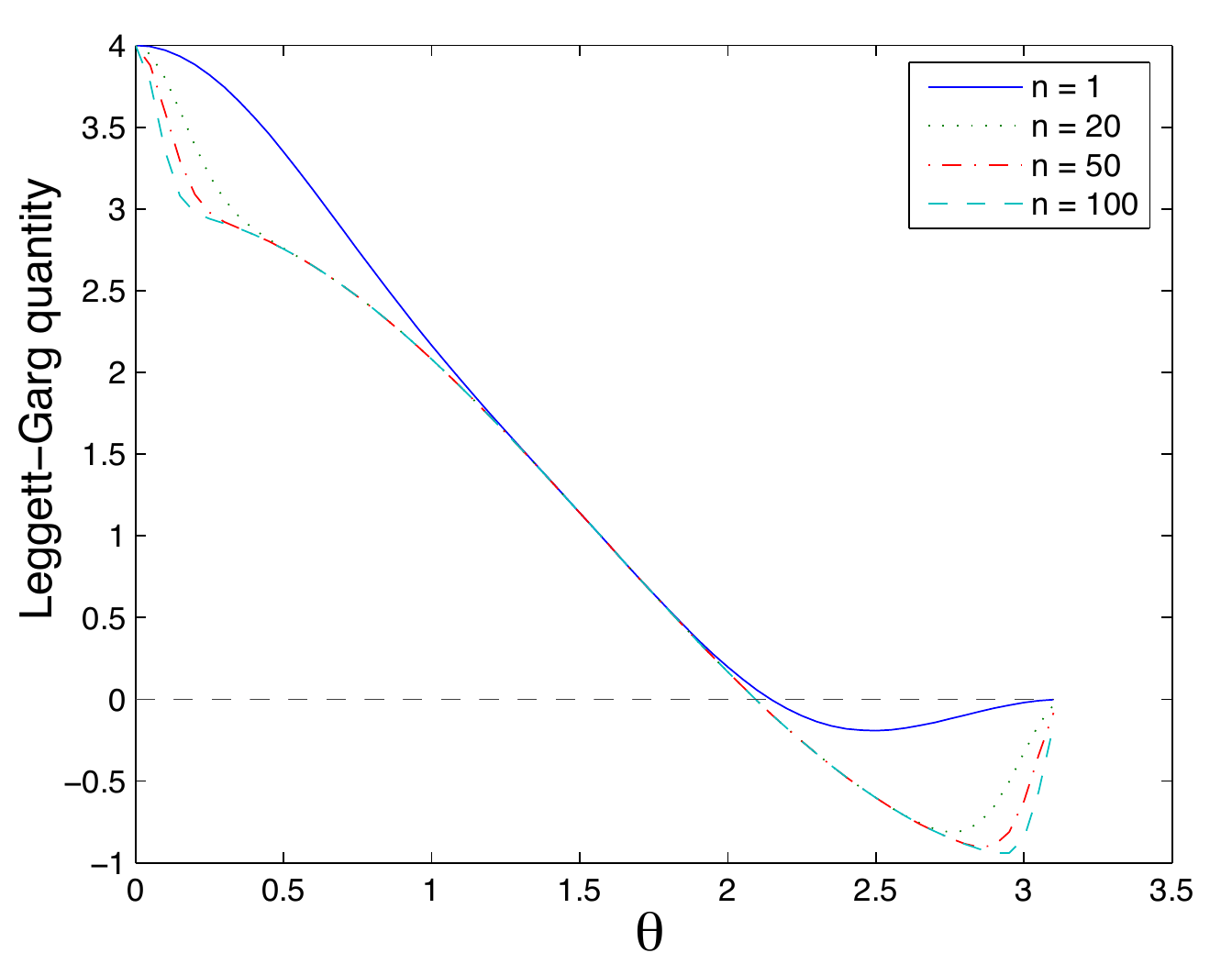}%
\caption{The above figure depicts the Leggett-Garg quantity in equation
(\ref{eq:LG-n}) for different values of the number $n$ of interleaved pairs of
measurements. The dashed line divides the space into two regions: points above
the line do not violate the Leggett-Garg inequality, while points below
violate it.}%
\label{fig:leggett-violation}%
\end{center}
\end{figure}


Let us suppose that the qubit begins in the maximally mixed state, so that
$\rho=I/2$. Then, the above expressions imply $\left\langle
Q_{1}Q_{3}\right\rangle ^\prime =\cos\left(  \theta\right)  $,\ $\left\langle
Q_{1}Q_{2}\right\rangle =\cos^{4}\left(  \theta\right)  $,\ $\left\langle
Q_{2}Q_{3}\right\rangle =\cos\left(  \theta\right)  $.  The Leggett-Garg
inequality
$
\mathcal{L}=1+\cos^{4}\left(  \theta\right)  +2\cos\left(  \theta\right)
\geq0
$
is violated if we choose $\theta$ between $.683\pi$ and $\pi$. If one could demonstrate this violation experimentally, it would show that our two-level quantum system is  {\it either} (i) non-macrorealistic {\it or} (ii) macrorealistic but with the peculiar property that two adroit measurements can somehow collude and strongly disturb the system.

\section{Generalization to more measurements}

We can generalize the
above analysis to the scenario where the experimentalist performs $2n+1$
measurements in the dotted box. We have $\sigma_{\theta}$ as $Q_{1}$; then $n$
interleaved pairs of measurements of $\sigma_{z}$ and $\sigma_{\theta}$
followed by $\sigma_{z}$; then $\sigma_{\theta}$ as $Q_{2}$; and finally
$\sigma_{z}$ as $Q_{3}$. In this case,
\begin{align*}
\left\langle Q_{1}Q_{2}\right\rangle  &  =\frac{1}{2}\text{Tr}\left[
\sigma_{z}\left(  \overline{\Delta}\circ\left(  \overline{\Delta}_{\theta
}\circ\overline{\Delta}\right)  ^{n}\right)  \left(  \left\{  \sigma_{\theta
},\rho\right\}  \right)  \right]  ,\\
\left\langle Q_{2}Q_{3}\right\rangle  &  =\frac{1}{2}\text{Tr}\left[
\sigma_{z}\left\{  \sigma_{\theta},\left(  \overline{\Delta}\circ
\overline{\Delta}_{\theta}\right)  ^{2n}\left(  \rho\right)  \right\}
\right]  ,
\end{align*}
while the other correlation function $\left\langle Q_{1}Q_{3}\right\rangle ^\prime$
remains the same. The measurements are again perfectly adroit in this ideal case.
If $\rho$ is maximally mixed, $\left\langle Q_{1}%
Q_{3}\right\rangle ^\prime=\cos\left(  \theta\right)  $,\ $\left\langle Q_{1}%
Q_{2}\right\rangle =\cos^{2\left(  n+1\right)  }\left(  \theta\right)
$,\ $\left\langle Q_{2}Q_{3}\right\rangle =\cos\left(  \theta\right)  $, then
the Leggett-Garg inequality reads as follows:%
\begin{equation}
\mathcal{L}=1+\cos^{2\left(  n+1\right)  }\left(  \theta\right)  +2\cos\left(
\theta\right)  \geq0. \label{eq:LG-n}%
\end{equation}
We can take the number $n$ of interleaved pairs of measurements to be
arbitrarily large. As shown in Fig.~\ref{fig:leggett-violation}, we obtain a
somewhat larger violation for a slightly larger range as $n$ increases
($2\pi/3\leq\theta < \pi$ when $n\rightarrow\infty$).  Note that it is not
possible to violate the inequality if one \emph{reduces} the number of
measurements in the dotted box in (e) from 3 to 1; although our protocol
clearly depends upon the non-commutativity of $\sigma_{z}$ and $\sigma
_{\theta}$, one does not obtain an inequality violation from the most
straightforward version of our protocol, where the measurements are
$\sigma_{\theta}$, then $\sigma_{z}$, then $\sigma_{\theta}$, then $\sigma
_{z}$. 

\section{Leggett-Garg Inequality with Non-zero Hamiltonian}
\label{sec:non-zero-Ham}
The original
Leggett-Garg inequality was framed for an rf-SQUID system with non-trivial dynamics. 
A violation of the inequality (\ref{eq:leggett-ineq}) occurs, for example, for a two-state quantum system such as an rf-SQUID initialized to a maximally mixed state with system Hamiltonian $\omega\sigma_{x}/2$, observable $Q$ chosen to be $\sigma_{z}$, and measurement times $t_{1}=0$, $t_{2}=3\pi/4\omega$, and $t_{3}=3\pi/2\omega$.  It is straightforward to adapt our protocol to this rf-SQUID case simply by converting
Fig.~\ref{leggett-experiment} from the Schr\"{o}dinger picture to the
Heisenberg picture. For simplicity, we take Hamiltonian $\omega\sigma_{x}$ so
that the evolution operator is  $U_t  \equiv\exp\left\{
-i\omega\sigma_{x}t\right\}  $, where we implicitly set $\hbar=1$. The
following relations hold%
\begin{align*}
\mathcal{U}_{t} \left( \sigma_{z} \right)   &  =\sigma_{y}%
\sin\left(  2\omega t\right)  +\sigma_{z}\cos\left(  2\omega t\right)  ,\\
\mathcal{U}_{t} \left( \sigma_{\theta} \right)   &  =\sigma
_{x}\sin\left(  \theta\right)  +\cos\left(  \theta\right) \left( \sigma_{y}\sin\left(  2\omega t\right)
    +\sigma_{z}\cos\left(  2\omega t\right) \right),
\end{align*}
where $\mathcal{U}_{t}\left(  \rho\right)  \equiv U_t  \rho
U^{\dag}_t$.
Notice that the second measurements in experiments (a)-(d) are all
perfectly adroit if we measure at time intervals equal to $\pi m/\omega$ where
$m$ is some positive integer. This adroitness holds because the function
$\sin\left(  2\omega t\right)$ vanishes at these times, and the scenario here then maps to the
earlier scenario with trivial dynamics. (In other words, these are quantum nondemolition measurements as in \cite{JKB06}.)  The correlation functions are as
follows:
\begin{align}
\left\langle Q_{1}Q_{3}\right\rangle ^\prime &  =\frac{1}{2}\text{Tr}\left[
Q_{3}\,\,\,\mathcal{U}_{5\tau}\left(\left\{  Q_{1},\mathcal{U}_{\tau}\left(
\rho\right)  \right\}  \right)  \right]  ,\label{eq:q1-q3-hamiltonian}\\
\left\langle Q_{1}Q_{2}\right\rangle  &  =\frac{1}{2}\text{Tr}\left[
Q_{2}\,\,\,\left(\mathcal{U}_{\tau}\circ\overline{\Delta}\circ \mathcal{U}_{\tau
}\circ \overline{\Delta}_{\theta}\circ \mathcal{U}_{\tau}\circ
\overline{\Delta}\circ  \mathcal{U}_{\tau}\right) \left(  \left\{  Q_{1}%
,\mathcal{U}_{\tau}\left(  \rho\right)  \right\}  \right) \right]  ,\label{eq:q1-q2-hamiltonian}\\
\left\langle Q_{2}Q_{3}\right\rangle  &  =\frac{1}{2}\text{Tr}\left[
Q_{3} \,\,\, \mathcal{U}_{\tau}\left(  \left\{  Q_{2},  \left(\mathcal{U}_{\tau}\circ
\overline{\Delta}\circ  \mathcal{U}_{\tau}\circ  \overline{\Delta}_{\theta
}\circ \mathcal{U}_{\tau}\circ  \overline{\Delta}\circ \mathcal{U}_{\tau
}\circ\overline{\Delta}_{\theta}\circ \mathcal{U}_{\tau}\right) \left(
\rho\right) \right\}  \right)  \right]  ,\label{eq:q2-q3-hamiltonian}
\end{align}
where $\tau$ is the uniform time interval between measurements. Choosing $\rho$ as the maximally mixed state,
$Q_{1}=\sigma_{\theta}$, $Q_{2}=\sigma_{\theta}$, $Q_{3}=\sigma_{z}$, and each
of the time intervals $\tau$ equal to $\pi m/\omega$ in equations
(\ref{eq:q1-q3-hamiltonian}-\ref{eq:q2-q3-hamiltonian}) yields the same
correlation values as in the trivial dynamics case. Thus, we obtain a
violation of the Leggett-Garg inequality if we measure at time intervals
$\tau$ equal to $\pi m/\omega$ where $m$ is some positive integer and if we
choose the angle $\theta$ to be in the range given previously.

\section{Leggett-Garg Inequality with Dephasing Noise}

\begin{figure}
[ptb]
\begin{center}
\includegraphics[
natheight=4.375100in,
natwidth=5.833200in,
width=4.0405in
]%
{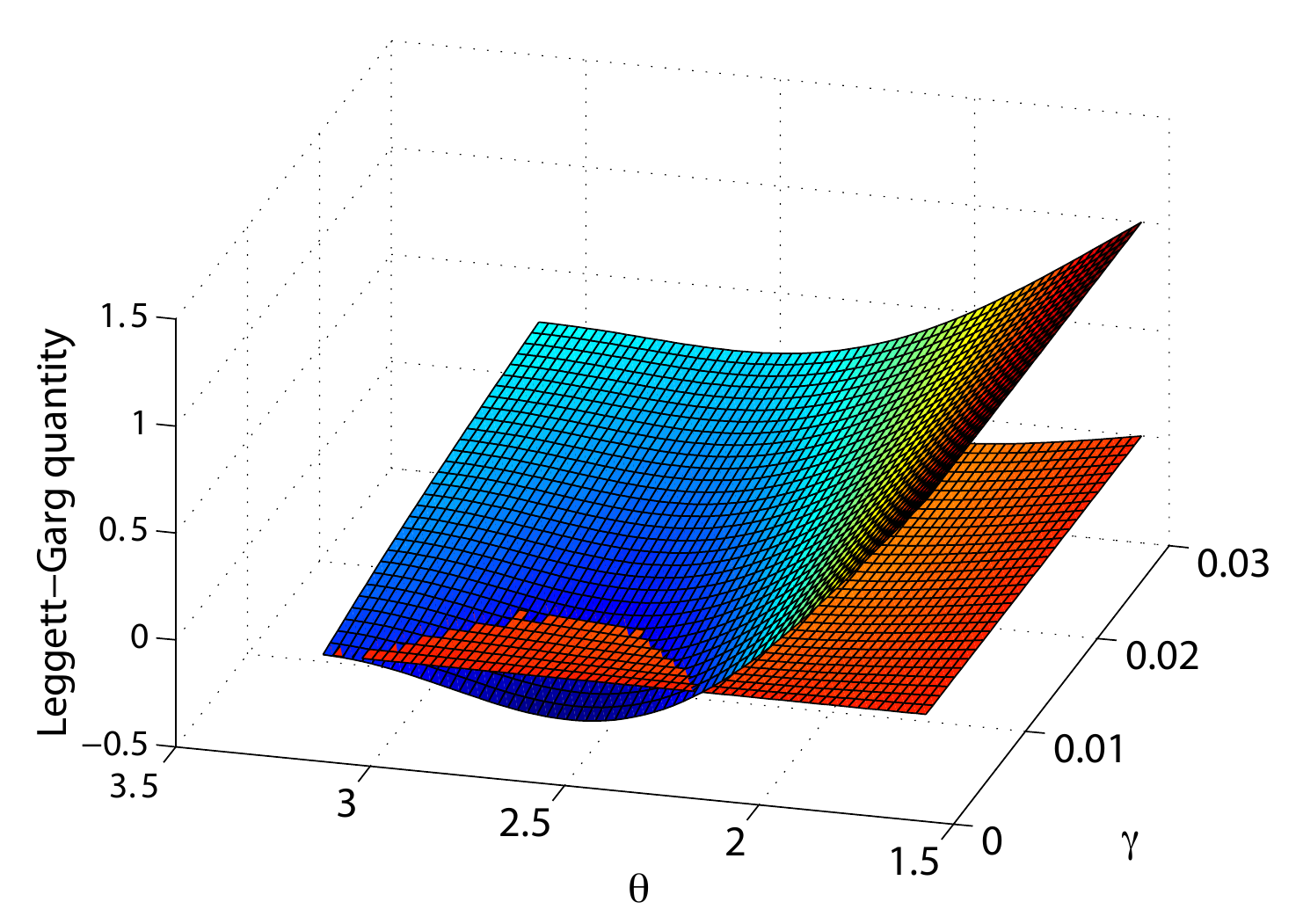}%
\caption{The red, flatter surface shows the function $-\epsilon_{\text{total}}(\gamma,\theta)$.  It results from
computing and summing the $\epsilon$-adroitness of
experiments (a)-(d) in Fig.~\ref{leggett-experiment}
when the system evolves according to the equation of
motion (\ref{eq:lindblad-dephasing}).
This surface divides the space into two regions:
points above the surface do not violate $\mathcal{L} \ge -\epsilon_{\text{total}}(\gamma,\theta)$, while points below the surface do violate it. The blue, curvier surface
shows the Leggett-Garg quantity $\mathcal{L}$ as a function of the angle $\theta$
and the dephasing rate $\gamma$. The range of angles $\theta$ for
which we observe a violation decreases as we increase the dephasing rate
$\gamma$. At $\gamma > 0.007$, it is no longer possible to observe a
violation for any angle $\theta$. The less stringent condition $\mathcal{L} \ge 0$ allows a violation up to $\gamma> 0.012$ (not shown).}%
\label{fig:leggett-dephase}%
\end{center}
\end{figure}

We modify the
above scenarios to include some dephasing effects described by the Lindblad
equation \cite{L76}:%
\begin{equation}
\dot{\rho}\left(  t\right)  =-i\left[  \omega\sigma_{x},\rho\left(  t\right)
\right]  +2\gamma\left(  \sigma_{z}\rho\left(  t\right)  \sigma_{z}%
-\rho\left(  t\right)  \right)  , \label{eq:lindblad-dephasing}%
\end{equation}
where the Hamiltonian is $\omega\sigma_{x}$ as before and $\gamma$ is the rate
of dephasing. Let $\mathcal{N}_{\tau}\left(  \rho\right)  $ denote the
time-dependent CPTP map that the above Lindblad equation effects. We can
calculate the correlation functions $\left\langle Q_{1}Q_{3}\right\rangle ^\prime$,
$\left\langle Q_{1}Q_{2}\right\rangle $, and $\left\langle Q_{2}%
Q_{3}\right\rangle $ by replacing $\mathcal{U}_{\tau}$ with $\mathcal{N}%
_{\tau}$ in equations (\ref{eq:q1-q3-hamiltonian}-\ref{eq:q2-q3-hamiltonian}).
We also compute the total amount of $\epsilon$-adroitness of the measurements in
experiments (a)-(d) of Fig.~\ref{leggett-experiment} using the definition of adroitness and change equation (\ref{eq:leggett-ineq}) to $\mathcal{L} \ge - \epsilon_{\text{total}}$.  This makes it harder to violate the Leggett-Garg inequality since only a substantially negative value of $\mathcal{L}$ is unambiguous given our finite measurement adroitness.  Fig.~\ref{fig:leggett-dephase}\ displays the Leggett-Garg quantity
$\mathcal{L}$\ as a function of the angle $\theta$ and the dephasing rate
$\gamma$. The range of angles $\theta$ for which we observe
a violation of the Leggett-Garg inequality decreases as we increase the
dephasing rate.%

One can also study Fig.~\ref{fig:leggett-violation} as a function of
dephasing rate $\gamma$. The violation exhibits a trade-off between the rate
$\gamma$ of dephasing noise and the number $n$ of pairs of interleaved
measurements because more measurements result in a longer amount of time for
dephasing, and this extra dephasing then offsets the benefits of more
measurements.

\section{Conclusion}

We have shown how to address a fundamental objection to
the Leggett-Garg inequality by altering the protocol. We then explored
improvements to our protocol and the effects of non-trivial system dynamics
and noise. Our generic Hamiltonian and dephasing model could be replaced with
more refined models, such as the spin-boson model \cite{RevModPhys.59.1}, for
calculations on specific systems. 

Note that only strong, projective measurements appear in our protocol.  Given the recent developments in weak-measurement-based Leggett-Garg tests \cite{GABLOWP09,PMNBVEK10,RKM06,JKB06,WJ08}, one might ask whether these tests have already addressed, or at least persuasively argued against, the clumsiness loophole.  Unfortunately, the answer is no.  The weak measurement-based Leggett-Garg tests are derived using a quantitative non-invasiveness axiom \cite{RKM06}  that assumes the detector noise $\xi\left(  t\right) $ and the system variable $Q\left(t+\tau\right)  $\  are uncorrelated in time: $\left\langle \xi\left(  t\right) Q\left(  t+\tau\right)  \right\rangle =0$.   The {\it reason} that a quantum system violates the weak measurement-based Leggett-Garg inequality is because it inevitably feels a sufficiently strong detector backaction $\left<\xi(t) Q(t+\tau)\right> = f(\tau) \ne 0$.  Since clumsy measurement of a macrorealistic system could also lead to strong detector backaction, a violation of the weak measurement-based Leggett-Garg inequality is perfectly consistent with the system being \textit{either} (i) non-macrorealistic \textit{or} (ii) macrorealistic but subjected to
a measurement technique that happens to disturb the system so that the
correlator $\left\langle \xi\left(  t\right)  Q\left(  t+\tau\right)
\right\rangle $ is non-zero.   The clumsiness loophole has been rephrased but not mitigated at all.

Nor is the clumsiness loophole mitigated by interesting observations like: the weaker the measurement, the larger the violation of the inequality \cite{GABLOWP09}.  A macrorealist cannot even make sense of such a claim without a definition of measurement strength.  The strength of a measurement is generally defined in terms of a quantum description of the system; a macrorealist does not accept the quantum description and would therefore require a definition of measurement strength in terms of some explicit experimental protocol.   Such a protocol, which has not yet been framed, would likely lead to loopholes at least as large as our collusion loophole.

In conclusion, we have presented a revision of the Leggett-Garg protocol that improves the rigor of tests of macrorealism.  This protocol should have applications both in quantum computing \cite{book2000mikeandike}%
\ and in the growing field of \textquotedblleft quantum biology\textquotedblright\ \cite{ADP08book,WMM09}.


\textit{Acknowledgements}---The authors thank \v{C}.~Brukner, T.~A.~Brun, A.~Cross, P.~Hayden,
A.~J.~Leggett, and J.~McCracken for useful discussions and acknowledge support
from SAIC, NSF, and the MDEIE (Qu\'{e}bec) PSR-SIIRI grant.

\end{document}